\documentclass[aps,prl,twocolumn,letter]{revtex4}
\usepackage{color}
\usepackage{graphicx}
\usepackage{dcolumn}
\usepackage{bm}

\begin{document}

\title{Spin-charge Separation in Nodal Antiferromagnetic Insulator }
\author{Su-Peng Kou}

  \affiliation{\ Department of Physics, Beijing Normal University, Beijing,
  100875 P. R. China }

\begin{abstract}
In this paper, by using two dimensional (2D) Hubbard models with
$\pi$-flux phase and that on a hexagonal lattice as examples, we
explore spin-charge-separated solitons in nodal antiferromagnetic
(AF) insulator - an AF order with massive Dirac fermionic
excitations (see detail in the paper). We calculate fermion zero
modes and induced quantum numbers on solitons (half skyrmions) in
the continuum limit, which are similar to that in the quasi
one-dimensional conductor polyacetylene \textrm{(CH)}$_x$ and that
in topological band insulator. In particular, we find some novel
phenomena : thanks to an induced staggered spin moment, a mobile
half skyrmion becomes a fermionic particle; when a hole or an
electron is added, the half skyrmion turns into a bosonic particle
with charge degree of freedom only. Our results imply that
nontrivial induced quantum number on solitons may be a universal
feature of spin-charge separation in different systems.

\end{abstract}

\maketitle

The Fermi liquid based view of the electronic properties has been very
successful as a basis for understanding the physics of conventional solids.
The quasi-particles of Fermi liquid carry both spin and charge quantum
numbers. However, in some cases, spin-charge separation occurs, providing a
new framework for thinking about the given systems. It indicates that the
systems have two independent elementary excitations, neutral spinon and
spinless holon, respectively, as opposed to single quasi-particle excitation
in conventional solids.

The first example is electronic systems in one spatial dimension\cite{le}.
The idea of solitons with induced quantum numbers starts with a beautiful
result obtained in the context of relativistic quantum field theories by
Jackiw and Rebbi\cite{jackiw}. Based on this idea, spin-charge separated
solitons had a lasting impact on condensed matter physics. In the long
molecule-chain of trans-Polyacetylene, spin-charge separation can occur in
term of soliton states\cite{polrev}. Due to induced fermion quantum numbers,
the soliton may be neutral particles with spin 1/2, or spinless with charge $%
\pm e$. In two dimensional electronic systems, spin-charge separation has
been a basic concept in understanding doped Mott-Hubbard insulator related
to high-$T_c$ cuprates \cite{pwa,KRS}. It is supposed that the particles can
be liberated at low energies, with spin-charge separation being an upshot in
the ``resonating valence bond''\ (RVB) spin liquid state which is proposed
by Anderson as a new state of matter\cite{pwa}. Recently topological band
insulator (TBI) has attracted considerable attention because of their
relevance to the quantum spin Hall effect\cite{km,sh}. It is pointed out
that there may exist spin-charge separated solitons in the presence of $\pi $
flux with induced quantum numbers\cite{qi,ran,ran1}.

In this paper we focus on a special class of antiferromagnetic (AF)
ordered state - nodal AF insulator, and we will show how spin-charge
separation occurs. Nodal AF insulator is an AF order (long range or
short range) with massive Dirac fermionic excitations. When there is
no AF order, fermionic excitations become nodal quasi-particles.
There are two examples of nodal AF insulator in condensed matter
physics - one is an AF order on a honeycomb lattice, the other is
$\pi $-flux phase together with a nonzero Neel order parameter.
Based on these examples, our results confirm that induced quantum
number on solitons is an important feature of the spin-charge
separation in different systems.

\label{meanf}\emph{Formulation}$\mathcal{-}$ To develop a systematical
formulation, we start with the extended Hubbard models,
\begin{equation}
H=-\sum\limits_{\langle i,j\rangle ,\sigma }t_{ij}\hat{c}_{i,\sigma
}^{\dagger }\hat{c}_{j,\sigma }+U\sum_j\hat{n}_{j\uparrow }\hat{n}%
_{j\downarrow }-\mu \sum\limits_{i,\sigma }\hat{c}_{i,\sigma }^{\dagger }%
\hat{c}_{i,\sigma }+h.c..
\end{equation}
Here $\hat{c}_{i,\sigma }^{\dagger }$ and $\hat{c}_{j,\sigma }$ are
electronic creation and annihilation operators. $U$ is the on-site Coulomb
repulsion. $\sigma $ are the spin-indices for electrons. $\mu $ is the
chemical potential. $\langle i,j\rangle $ denotes two sites on a
nearest-neighbor link. $\hat{n}_{j\uparrow }$ and $\hat{n}_{j\downarrow }$
are the number operators of electrons with up-spin and down spin. On a
honeycomb lattice, the nearest neighbor hopping is a constant, $t_{i,j}=t;$
on a square lattice with $\pi $-flux phase, it can be chosen as $t_{i,i+\hat{%
x}}=\chi ,$ $t_{i,i+\hat{y}}=i\chi $ \cite{st,st0,pi,wen}. The partition
function of the extended Hubbard models is written as $\mathcal{Z}=\int
\mathcal{D}\overline{c}\mathcal{D}ce^{-\int_0^\beta d\tau L},$ where
\begin{eqnarray}
L &=&\sum_{j,\sigma }\bar{c}_{j,\sigma }\left( \partial _\tau -\mu \right)
c_{j,\sigma }+\sum_{\langle i,j\rangle ,\sigma }t_{ij}\bar{c}_{i,\sigma
}c_{j,\sigma } \\
&&-U\sum_jn_{j\uparrow }n_{j\downarrow }.  \nonumber
\end{eqnarray}
$\bar{c}_{i,\sigma }$ and $c_{j,\sigma }$ are Grassmann variables describing
the electronic fields.

Firstly let us derive long wave-length effective Lagrangian of the hopping
term in the extended Hubbard models. Although $\pi $-flux phase does not
break translational symmetry, we may still divide the square lattice into
two sublattices, $A$ and $B$. After transforming the hopping term into
momentum space, we obtain $E_f=2\chi \sqrt{\cos ^2k_x+\cos ^2k_y}$. So there
exist two nodal fermi-points at $\mathbf{k}_1=(\frac \pi 2,\frac \pi 2),$ $%
\mathbf{k}_2=(\frac \pi 2,-\frac \pi 2)$ and the spectrum of fermions
becomes linear in the vicinity of the two nodal points. On a honeycomb
lattice, after dividing the lattice into two sublattices, $A$ and $B$, the
dispersion is obtained in Ref.\cite{her,her1,pei}. There also exist two
nodal points, $\mathbf{k}_1=\frac{2\pi }{\sqrt{3}}(1,\frac 1{\sqrt{3}})$ and
$\mathbf{k}_2=\frac{2\pi }{\sqrt{3}}(-1,-\frac 1{\sqrt{3}})$ and the
spectrum of fermions becomes linear near $\mathbf{k}_{1,2}$. In the
continuum limit, the Dirac-like effective Lagrangian describes the low
energy fermionic modes for both cases
\begin{equation}
\mathcal{L}_f=i\bar{\psi}_1\gamma _\mu \partial _\mu \psi _1+i\bar{\psi}%
_2\gamma _\mu \partial _\mu \psi _2
\end{equation}
where $\bar{\psi}_1=\psi _1^{\dagger }\gamma _0=(
\begin{array}{llll}
\bar{\psi}_{\uparrow 1A}, & \bar{\psi}_{\uparrow 1B}, & \bar{\psi}%
_{\downarrow 1A}, & \bar{\psi}_{\downarrow 1B}
\end{array}
)$ and $\bar{\psi}_2=\psi _2^{\dagger }\gamma _0=(
\begin{array}{llll}
\bar{\psi}_{\uparrow 2B}, & \bar{\psi}_{\uparrow 2A}, & \bar{\psi}%
_{\downarrow 2B}, & \bar{\psi}_{\downarrow 2A}
\end{array}
)$\cite{her,her1,pei}$.$ $\gamma _\mu $ is defined as $\gamma _0=\sigma
_0\otimes \tau _z,$ $\gamma _1=\sigma _0\otimes \tau _y,$ $\gamma _2=\sigma
_0\otimes \tau _x,$ $\sigma _0=\left(
\begin{array}{ll}
1 & 0 \\
0 & 1
\end{array}
\right) $. $\tau ^x,$ $\tau ^y,$ $\tau ^z$ are Pauli matrices. We have set
the Fermi velocity to be unit $v_F=1$.

In the strongly coupling limit, $U>>t$, there always exists an AF order in
the extended Hubbard models. Introducing Stratonovich-Hubbard fields for the
spin degrees of freedom \cite{wen}, we obtain the partition function as $%
Z=\int \mathcal{D}\overline{c}\mathcal{D}c\mathcal{D}\mathbf{B}%
e^{-\int_0^\beta d\tau L}$, where the Lagrangian is given by
\begin{eqnarray}
L &=&\sum_{j,\sigma }\bar{c}_{j,\sigma }\left( \partial _\tau -\mu \right)
c_{j,\sigma }+\sum_{\langle i,j\rangle ,\sigma }t_{ij}\bar{c}_{i,\sigma
}c_{j,\sigma }  \label{mo} \\
&&-\frac 3{2U}\sum_j\mathbf{B}_j^2+U\sum_j(-1)^j\mathbf{B}_j\cdot \bar{c}_j%
\mathbf{\sigma }c_j  \nonumber
\end{eqnarray}
with Pauli matrices $\mathbf{\sigma =}(\sigma ^x,\sigma ^y,\sigma ^z)\mathbf{%
.}$ Here $\mathbf{B}_j$ is a vector denoting spin configurations, $\mathbf{B}%
_j=\left| B_j\right| \mathbf{n}_j$ where $\left| B_j\right| =\phi _0$
represents the value of localized spin moments and $\mathbf{n}_j$ is a unit
vector describing the N\'{e}el order parameter. In AF ordered state, the
mass gap of the electrons is given as $m=\phi _0U$.

Then starting from Eq.(\ref{mo}), we get the same long wave-length
effective model of nodal AF insulator \cite{at,ki,senthil}
\begin{equation}
\mathcal{L}_{\mathrm{eff}}=\sum\limits_\alpha i\bar{\psi}_\alpha \gamma _\mu
\partial _\mu \psi _\alpha +m(\bar{\psi}_1\mathbf{n\cdot \sigma }\psi _1-%
\bar{\psi}_2\mathbf{n\cdot \sigma }\psi _2).  \label{mo1}
\end{equation}
$\alpha =1,$ $2$ labels the two Fermi points

\emph{Zero modes on half skyrmions} - In this section we will study the
properties of topological solitons. Instead of considering topological
solitons with integer topological charge (skyrmions), we focus on solitons
with a half topological charge, $\int \frac 1{2\pi }\mathbf{n}\cdot \partial
_x\mathbf{n}\times \partial _y\mathbf{n}$ $d^2\mathbf{r}=\pm \frac 12$ with $%
\mathbf{r=(}x,$ $y\mathbf{).}$ Such soliton is called a half skyrmion (meron
or anti-meron). A meron with a narrow core size (the lattice size $a$) is
characterized by $\mathbf{n=r}/\mid \mathbf{r}\mid ,$\ $\mathbf{r}^2=x^2+y^2$
\cite{pol,meron,half,ba,ng,ots,weng1,kou}. To stabilize such a solitons, one
may add a small easy-plane anisotropy of the N\'{e}el order.

Around a meron configuration, the fermionic operators are expanded as
\begin{eqnarray}
\hat{\psi}_{\mathbf{\alpha }}(\mathbf{r},t) &=&\sum_{\mathbf{k}\neq 0}\hat{b}%
_{\mathbf{\alpha k}}e^{-iE_{\mathbf{k}}t}\psi _{{\mathbf{\alpha }}\mathbf{k}%
}(\mathbf{r}) \\
&&+\sum_{\mathbf{k}\neq 0}\hat{d}_{\mathbf{\alpha k}}^{\dagger }e^{iE_{%
\mathbf{k}}t}\psi _{\mathbf{\alpha k}}^{\dagger }(\mathbf{r})+\hat{a}_{%
\mathbf{\alpha }}^0\psi _{\mathbf{\alpha }}^0(\mathbf{r}),\text{ }  \nonumber
\end{eqnarray}
where $\hat{b}_{\alpha \mathbf{k}}$ and $\hat{d}_{\alpha \mathbf{k}%
}^{\dagger }$ are operators of $\mathbf{k}\neq 0$ modes that are irrelevant
to the soliton states discussed below. $\psi _{\mathbf{\alpha k}}^{\dagger }(%
\mathbf{r})=(
\begin{array}{llll}
\psi _{\uparrow \mathbf{\alpha Ak}}^{0*}, & \psi _{\uparrow \mathbf{\alpha }B%
\mathbf{k}}^{0*}, & \psi _{\downarrow \mathbf{\alpha Ak}}^{0*}, & \psi
_{\downarrow \mathbf{\alpha }B\mathbf{k}}^{0*}
\end{array}
)$ are the functions of zero modes. $\hat{a}_\alpha ^0$ are annihilation
operators of zero modes.

To obtain the zero modes, we write down two Dirac equations from Eq.(\ref
{mo1})
\begin{equation}
i\partial _x\gamma _1\psi _1^0+i\partial _y\gamma _2\psi _1^0+m\mathbf{%
n\cdot \sigma }\psi _1^0=0
\end{equation}
and
\begin{equation}
i\partial _x\gamma _1\psi _2^0+i\partial _y\gamma _2\psi _2^0-m\mathbf{%
n\cdot \sigma }\psi _2^0=0.
\end{equation}
Firstly we solve the Dirac equation for $\psi _1^0$. With the ansatz $\psi
_1^0=\left(
\begin{array}{l}
\xi _1(\tilde{x})e^{-i\theta } \\
\xi _2(\tilde{x}) \\
\xi _3(\tilde{x}) \\
\xi _4(\tilde{x})e^{i\theta }
\end{array}
\right) ,$ we have
\begin{eqnarray}
\partial _{\tilde{x}}\xi _2 &=&\xi _3,\text{ }\partial _{\tilde{x}}\xi
_3=\xi _2, \\
\partial _{\tilde{x}}\xi _1 &=&-\frac{\xi _1}{\tilde{x}}+\xi _4,\text{ }%
\partial _{\tilde{x}}\xi _4=-\frac{\xi _4}{\tilde{x}}+\xi _1  \nonumber
\end{eqnarray}
where $\mathbf{r=}\mid \mathbf{r}\mid (\cos \theta ,\sin \theta )$ and $%
\tilde{x}=\frac{\mid \mathbf{r}\mid }m$. The solution has been obtained in
Ref.\cite{zeromode} as
\begin{equation}
\xi _1(\tilde{x})=\xi _4(\tilde{x})=0,\text{ }\xi _2(\tilde{x})=\xi _3(%
\tilde{x})=\tilde{x}^{\frac 12}K_{\frac 12}(\tilde{x})  \nonumber
\end{equation}
where $K_{\frac 12}(\tilde{x})$ is the modified Bessel function. So the
solution of $\psi _1^0$ becomes $\left(
\begin{array}{l}
0 \\
\tilde{x}^{\frac 12}K_{\frac 12}(\tilde{x}) \\
\tilde{x}^{\frac 12}K_{\frac 12}(\tilde{x}) \\
0
\end{array}
\right) .$

To solve $\psi _2^0,$ we transform the equation $i\partial _i\gamma _i\psi
_2^0-m\mathbf{n\cdot \sigma }\psi _2^0=0$ into
\begin{equation}
Ui\partial _i\gamma _i\tilde{\psi}_2^0U^{-1}+mU\mathbf{n\cdot \sigma }\tilde{%
\psi}_2^0U^{-1}=0,
\end{equation}
where $U=e^{i\pi \gamma _0/2},$ $U\gamma _iU^{-1}=-\gamma _i$ and $%
U^{-1}\psi _2^0U=\tilde{\psi}_2^0.$ Then the solution of $\psi _2^0\ $is
obtained as $\left(
\begin{array}{l}
0 \\
-\tilde{x}^{\frac 12}K_{\frac 12}(\tilde{x}) \\
\tilde{x}^{\frac 12}K_{\frac 12}(\tilde{x}) \\
0
\end{array}
\right) .$

It is noticable that from above solutions of zero modes, the components $%
\psi _{\uparrow 1A}^0,$ $\psi _{\downarrow 1B}^0,$ $\psi _{\downarrow 2A}^0$
and $\psi _{\uparrow 2B}^0$ are all zero.

\emph{Topological mechanism of spin-charge separation - } For the solutions
of zero modes, there are four zero-energy soliton states $\mid \mathrm{sol}%
\rangle $ around a half skyrmion which are denoted by $\mid 1_{+}\rangle
\otimes \mid 2_{+}\rangle ,$ $\mid 1_{-}\rangle \otimes \mid 2_{-}\rangle ,$
$\mid 1_{-}\rangle \otimes \mid 2_{+}\rangle $ and $\mid 1_{+}\rangle
\otimes \mid 2_{-}\rangle .$ $\mid 1_{-}\rangle $ and $\mid 2_{-}\rangle $
are empty states of the zero modes $\psi _1^0(\mathbf{r}) $ and $\psi _2^0(%
\mathbf{r});$ $\mid 1_{+}\rangle $ and $\mid 2_{+}\rangle $ are occupied
states of them. Thus we have the relationship between $\hat{a}_{\mathbf{%
\alpha }}^0$ and $\mid \mathrm{sol}\rangle $ as
\begin{equation}
\hat{a}_1^0\mid 1_{+}\rangle =\mid 1_{-}\rangle ,\text{ }\hat{a}_1^0\mid
1_{-}\rangle =0,\text{ }\hat{a}_2^0\mid 2_{+}\rangle =\mid 2_{-}\rangle ,%
\text{ }\hat{a}_2^0\mid 2_{-}\rangle =0,  \label{a}
\end{equation}

Firstly we define total induced fermion number operators of the soliton
states, $\hat{N}_F=\sum\limits_\alpha \hat{N}_{\alpha ,F}$ with
\begin{eqnarray}
\hat{N}_{\alpha ,F} &\equiv &\int :\hat{\psi}_\alpha ^{\dagger }\hat{\psi}%
_\alpha :d^2\mathbf{r} \\
&=&(\hat{a}_\alpha ^0)^{\dagger }\hat{a}_\alpha ^0+\sum\limits_{\mathbf{k}%
\neq 0}(\hat{b}_{\alpha \mathbf{k}}^{\dagger }\hat{b}_{\alpha \mathbf{k}}-%
\hat{d}_{\alpha \mathbf{k}}^{\dagger }\hat{d}_{\alpha \mathbf{k}})-\frac 12.
\nonumber
\end{eqnarray}
$:\hat{\psi}_\alpha ^{\dagger }\hat{\psi}_\alpha :$ means normal product of $%
\hat{\psi}_\alpha ^{\dagger }\hat{\psi}_\alpha $. From the relation between $%
\hat{a}_{\mathbf{\alpha }}^0$ and $\mid \mathrm{sol}\rangle $ in Eq.(\ref{a}%
), we find that $\mid 1_{\pm }\rangle $ or $\mid 2_{\pm }\rangle $ have
eigenvalues of $\pm \frac 12$ of the total induced fermion number operator $%
\hat{N}_F $,
\begin{eqnarray}
\hat{N}_{1,F}|1_{\pm }\rangle &=&\pm {\frac 12}|1_{\pm }\rangle ,\text{ }%
\hat{N}_{1,F}|2_{\pm }\rangle =0,\text{ }  \label{half} \\
\hat{N}_{2,F}|2_{\pm }\rangle &=&\pm {\frac 12}|2_{\pm }\rangle ,\text{ }%
\hat{N}_{2,F}|1_{\pm }\rangle =0.  \nonumber
\end{eqnarray}

Another important induced quantum number operator is staggered spin
operator, $\hat{S}_{(\pi ,\pi )}^z=\frac 12\sum\limits_{i\in A}\hat{c}%
_i^{\dagger }\sigma _z\hat{c}_i-\frac 12\sum\limits_{i\in B}\hat{c}%
_i^{\dagger }\sigma _z\hat{c}_i=\frac 12\int :[(\hat{\psi}_{\uparrow
1A}^{\dagger }\hat{\psi}_{\uparrow 1A}+\hat{\psi}_{\downarrow 1B}^{\dagger }%
\hat{\psi}_{\downarrow 1B}-\hat{\psi}_{\downarrow 1A}^{\dagger }\hat{\psi}%
_{\downarrow 1A}$ $-\hat{\psi}_{\uparrow 1B}^{\dagger }\hat{\psi}_{\uparrow
1B})+(\hat{\psi}_{\uparrow 2A}^{\dagger }\hat{\psi}_{\uparrow 2A}+\hat{\psi}%
_{\downarrow 2B}^{\dagger }\hat{\psi}_{\downarrow 2B}-\hat{\psi}_{\downarrow
2A}^{\dagger }\hat{\psi}_{\downarrow 2A}-\hat{\psi}_{\uparrow 2B}^{\dagger }%
\hat{\psi}_{\uparrow 2B})]:d^2\mathbf{r}.$ For the four degenerate zero
modes, it can be simplified into $\hat{S}_{(\pi ,\pi )}^z\mid \mathrm{sol}%
\rangle =\frac 12(\hat{N}_{2,F}-\hat{N}_{1,F})\mid \mathrm{sol}\rangle .$
Let us show the detailed calculations. From the zero solutions of $\psi
_{\uparrow 1A}^0,$ $\psi _{\downarrow 1B}^0,$ $\psi _{\downarrow 2A}^0$ and $%
\psi _{\uparrow 2B}^0$, we obtain four equations
\begin{eqnarray}
(\int &:&\hat{\psi}_{\uparrow 1A}^{\dagger }\hat{\psi}_{\uparrow 1A}:d^2%
\mathbf{r})\mid \mathrm{sol}\rangle \equiv 0,  \label{ze} \\
(\int &:&\hat{\psi}_{\downarrow 1B}^{\dagger }\hat{\psi}_{\downarrow 1B}:d^2%
\mathbf{r})\mid \mathrm{sol}\rangle \equiv 0,  \nonumber \\
(\int &:&\hat{\psi}_{\downarrow 2A}^{\dagger }\hat{\psi}_{\downarrow 2A}:d^2%
\mathbf{r})\mid \mathrm{sol}\rangle \equiv 0,  \nonumber \\
(\int &:&\hat{\psi}_{\uparrow 2B}^{\dagger }\hat{\psi}_{\uparrow 2B}:d^2%
\mathbf{r})\mid \mathrm{sol}\rangle \equiv 0.  \nonumber
\end{eqnarray}
Using above four equations, we obtain
\begin{widetext}
\[
\hat{S}_{(\pi ,\pi )}^z\mid \mathrm{sol}\rangle =\frac 12\int d^2\mathbf{r}%
:(-\hat{\psi}_{\uparrow 1A}^{\dagger }\hat{\psi}_{\uparrow 1A}-\hat{\psi}%
_{\downarrow 1B}^{\dagger }\hat{\psi}_{\downarrow 1B}-\hat{\psi}_{\downarrow
1A}^{\dagger }\hat{\psi}_{\downarrow 1A}-\hat{\psi}_{\uparrow 1B}^{\dagger }%
\hat{\psi}_{\uparrow 1B}
\]
\[
+\hat{\psi}_{\uparrow 2A}^{\dagger }\hat{\psi}_{\uparrow 2A}+\hat{\psi}%
_{\downarrow 2B}^{\dagger }\hat{\psi}_{\downarrow 2B}+\hat{\psi}_{\downarrow
2A}^{\dagger }\hat{\psi}_{\downarrow 2A}+\hat{\psi}_{\uparrow 2B}^{\dagger }%
\hat{\psi}_{\uparrow 2B}):\mid \mathrm{sol}\rangle =-\frac 12(\hat{N}_{1,F}-%
\hat{N}_{2,F})\mid \mathrm{sol}\rangle .
\]\end{widetext}

Then we calculate two induced quantum numbers defined above. Without doping,
the soliton states of a half skyrmion are denoted by $\mid 1_{-}\rangle
\otimes \mid 2_{+}\rangle $ and $\mid 1_{+}\rangle \otimes \mid 2_{-}\rangle
$. One can easily check that the total induced fermion number on the
solitons is zero from the cancelation effect between two nodals $\hat{N}%
_F\mid 1_{-}\rangle \otimes \mid 2_{+}\rangle =\hat{N}_F\mid 1_{+}\rangle
\otimes \mid 2_{-}\rangle =0.$ It is consistent to the earlier results that
forbid a Hopf term for the low energy theory of two dimensional Heisenberg
model\cite{wenhopf}. On the other hand, there exists \textit{an}\textit{\
induced staggered spin moment} on the soliton states $\mid 1_{-}\rangle
\otimes \mid 2_{+}\rangle $ and $\mid 1_{+}\rangle \otimes \mid 2_{-}\rangle
,$
\begin{eqnarray}
\hat{S}_{(\pi ,\pi )}^z &\mid &1_{-}\rangle \otimes \mid 2_{+}\rangle =\frac
12\mid 1_{-}\rangle \otimes \mid 2_{+}\rangle , \\
\hat{S}_{(\pi ,\pi )}^z &\mid &1_{+}\rangle \otimes \mid 2_{-}\rangle
=-\frac 12\mid 1_{+}\rangle \otimes \mid 2_{-}\rangle .  \nonumber
\end{eqnarray}
The induced staggered spin moment may be straightforwardly obtained by
combining the definition of $\hat{S}_{(\pi ,\pi )}^z$ and Eq.(\ref{half})
together.

When half skyrmions become mobile, their quantum statistics becomes
important. Let us examine the statistics of a half skyrmion with an induced
staggered spin moment. In \textrm{CP(1)} representation of $\mathbf{n,}$ a
''bosonic spinon'' is introduced by $\mathbf{n}=\mathbf{\bar{z}\sigma z}$
with $\mathbf{z}=\left(
\begin{array}{l}
z_{\uparrow } \\
z_{\downarrow }
\end{array}
\right) $ and $\mathbf{\bar{z}z=1}$. Since each ''bosonic spinon'' $\mathbf{z%
}$ carries $\pm \frac 12$ staggered spin moment, an induced staggered spin
moment corresponds to a trapped "bosonic spinon" $\mathbf{z}$. On the other
hand, a half skyrmion can be regarded as a $\pi -$flux of the "bosonic
spinon", $\int \frac 1{2\pi }\mathbf{n}\cdot \partial _x\mathbf{n}\times
\partial _y\mathbf{n}$ $d^2\mathbf{r}=\frac 1{2\pi }\int \epsilon _{\mu \nu
}\partial _\mu a_\nu d^2\mathbf{r}=\pm \frac 12$ with $a_\mu \equiv \frac i2(%
\mathbf{\bar{z}}\partial _\mu \mathbf{z}-\partial _\mu \mathbf{\bar{z}z}).$
To be more explicit, moving a ''bosonic spinon'' $z$ around a half skyrmion
generates a Berry phase $\phi $ to $\mathbf{z}\rightarrow \mathbf{z}^{\prime
}=\left(
\begin{array}{l}
z_{\uparrow }e^{i\phi } \\
z_{\downarrow }e^{i\phi }
\end{array}
\right) $ where $\phi =\int \epsilon _{\mu \nu }\partial _\mu a_\nu d^2%
\mathbf{r}=\pm \pi $. As a result, a ''bosonic spinon'' $\mathbf{z}$ and a
half skyrmion (meron or antimeron ) share mutual semion statistics. Binding
the trapped ''bosonic spinon'', a mobile half skyrmion becomes \textit{a
fermionic particle}. We may use the operator $\hat{f}_\sigma $ to describe
such neutral fermionic particle with half spin. The relation between the
zero energy states and the fermionic states is given as $\mid 1_{+}\rangle
\otimes \mid 2_{-}\rangle =\hat{f}_{\downarrow }^{\dagger }\mid 0\rangle _f%
\text{ and }\mid 1_{-}\rangle \otimes \mid 2_{+}\rangle =\hat{f}_{\uparrow
}^{\dagger }\mid 0\rangle _f$ (The state $|0\rangle _f$ is defined through $%
\hat{f}_{\uparrow }|0\rangle _f=\hat{f}_{\downarrow }|0\rangle _f=0$). We
call such neutral object (fermion with $\pm \frac 12$ spin degree freedom) a
(fermionic) ''spinon''.

Now we go away from half filling. It is known that when a hole (electron) is
doped, it is equivalence to removing (adding) an electron. Without
considering the existence of half skyrmions, the hole (electron) will be
doped into the lower (upper) Hubbard band. The existence of zero modes on
half skyrmions leads to the appearance of bound levels in the middle of the
Mott-Hubbard gap \cite{meron}. The hole (electron) will be doped onto the
bound states on the half skyrmion and then one of the zero modes is
occupied. When one hole is doped, the soliton state is denoted by $%
|1_{-}\rangle \otimes |2_{-}\rangle $. One can easily check the result by
calculating its induce quantum numbers. On the one hand, there is no induced
staggered spin moment, $\hat{S}_{(\pi ,\pi )}^z\mid 1_{-}\rangle \otimes
\mid 2_{-}\rangle =0.$ On the other hand, the total fermion number is not
zero, $\hat{N}_F|1_{-}\rangle \otimes |2_{-}\rangle =-|1_{-}\rangle \otimes
|2_{-}\rangle .$ These results mean that such soliton state is a spinless
''holon'' with positive charge degrees of freedom. After binding a fermionic
hole, the soliton state (holon) does not have an induced staggered spin
moment. Thus the holon obeys bosonic statistics and becomes a \textit{%
charged bosonic particles}. When one electron is doped, the soliton state is
denoted by $|1_{+}\rangle \otimes |2_{+}\rangle $. The induced quantum
numbers of it are $\hat{N}_F|1_{+}\rangle \otimes |2_{+}\rangle
=+|1_{+}\rangle \otimes |2_{+}\rangle $ and $\hat{S}_{(\pi ,\pi )}^z\mid
1_{+}\rangle \otimes \mid 2_{+}\rangle =0.$ Such soliton state is also a
bosonic particle with a negative charge but without spin degrees of freedom.
We call such a soliton state an ''electon'' to mark difference with the word
''electron''.

      \begin{table}[t]
      \begin{tabular}{|c|cccc|}
      \hline $$ & $\mid 1_{+}\rangle \otimes \mid 2_{+}\rangle$ & $\mid
      1_{-}\rangle \otimes \mid 2_{+}\rangle$ & $\mid 1_{+}\rangle \otimes
      \mid 2_{-}\rangle $ & $\mid 1_{-}\rangle \otimes \mid 2_{-}\rangle$
      \\ \hline
      $N_F$ & $1$ & 0 & 0 & -1 \\
      $S_{(\pi ,\pi )}^z$ & $0$ & $1/2$ & $-1/2$ & 0 \\ \hline
      \end{tabular}
      \caption{Quantum numbers of the degenerate soliton states.}
      \label{Z2Ecm}
      \end{table}

Finally we get a\emph{\ }\textit{topological mechanism of spin-charge
separation} in nodal AF insulators. There exist two types of topological
objects - one is the fermionic spinon, the other is the bosonic holon ( or
the bosonic electon ).

In 1D system, real spin-charge separation may occur. As far as the low
energy physics is concerned, the spin and charge dynamics are completely
decoupled from each other. In 2D, real spin-charge separation in a nodal AF
insulators can not occur in long range AF order. In the future we will study
the deconfinement condition of spin-charge separated solitons and explore
the properties of deconfined phases with real spin-charge separation.

\emph{Summary} - By using 2D $\pi $-flux phase Hubbard model and the Hubbard
model on a honeycomb lattice as examples, we explore spin-charge separation
in nodal AF insulator. The crus crux of the matter in this paper is the
discovery of induced staggered spin moment $S_{(\pi ,\pi )}^z$ on half
skyrmions in nodal AF insulators. Based on such nontrivial induced quantum
number, we classify four degenerate soliton states with zero energy - two of
them ($\mid 1_{-}\rangle \otimes \mid 2_{+}\rangle $ and $\mid 1_{+}\rangle
\otimes \mid 2_{-}\rangle $) represent the up-spin and down-spin states for
a fermionic ''spinon'', another state ($|1_{-}\rangle \otimes |2_{-}\rangle $%
) represents a ''holon'' and the last one ($|1_{+}\rangle \otimes
|2_{+}\rangle $) denotes an ''electon''.

This research is supported by NFSC Grant no. 10574014.

\end{document}